\documentclass[a4paper,11pt]{article}
\usepackage{jcappub}
\usepackage{aas_macros}
\usepackage{subcaption}

\newcommand{\mumin}{\mu_\mathrm{min}}
\newcommand{\mumax}{\mu_\mathrm{max}}
\newcommand{\Nran}{N_\mathrm{ran}}
\newcommand{\Ngal}{N_\mathrm{gal}}

\title{Imprint of DESI fiber assignment on the anisotropic power 
spectrum of emission line galaxies}
\author[a,d]{Lucas Pinol,}
\author[d]{Robert N. Cahn,}
\author[c,d]{Nick Hand,}
\author[b,c,d]{Uro\v s Seljak,}
\author[b,c]{Martin White}

\affiliation[a]{D\'epartement de Physique, \'Ecole Normale Sup\'erieure, Paris, France}
\affiliation[b]{Department of Physics, University of California, Berkeley, California}
\affiliation[c]{Department of Astronomy, University of California, Berkeley, California}
\affiliation[d]{Lawrence Berkeley National Laboratory, Berkeley, California}

\abstract{
The Dark Energy Spectroscopic Instrument (DESI), a multiplexed 
fiber-fed spectrograph, is a Stage-IV ground-based dark energy experiment aiming to 
measure redshifts for 29 million Emission-Line Galaxies (ELG), 4 million Luminous Red 
Galaxies (LRG), and 2 million Quasi-Stellar Objects (QSO). 
The survey design includes a pattern of tiling on the sky and the locations of the 
fiber positioners in the focal plane of the telescope, with the observation strategy
determined by a fiber assignment algorithm that optimizes the allocation of fibers to 
targets. This strategy allows a given region to be covered on average five times for 
a five-year survey, but with coverage varying between zero and twelve, which imprints 
a spatially-dependent pattern on the galaxy clustering. We investigate the systematic
effects of the fiber assignment coverage on the anisotropic galaxy clustering of 
ELGs and show that, in the absence of any corrections, it leads to discrepancies of 
order ten percent on large scales for the power spectrum multipoles. We introduce a 
method where objects in a random catalog are assigned a coverage, and the mean density is 
separately computed for each coverage factor. We show that this method reduces, 
but does not eliminate the effect. We next investigate the angular dependence of the
contaminated signal, arguing that it is mostly localized to purely transverse modes. 
We demonstrate that the cleanest way to remove the contaminating signal is to perform 
an analysis of the anisotropic power spectrum $P(k,\mu)$ and remove the lowest 
$\mu$ bin, leaving $\mu>0$ modes accurate at the few-percent level. Here, $\mu$ 
is the cosine of the angle between the line-of-sight and the direction of $\vec{k}$.
We also investigate two alternative definitions of the random catalog and 
show they are comparable but less effective than the coverage randoms method.}

\begin{document}
\maketitle

\section{Introduction}

The Dark Energy Spectroscopic Instrument (DESI) is a 5000-fiber spectroscopic 
instrument, which will enable massively parallel measurements of galaxy 
and quasar redshifts \cite{Levi2013,DESIInstrumentDesign:2016,DESIScienceDesign:2016}.
Installed at the Mayall 4-meter telescope 
at Kitt Peak, Arizona, DESI will make the largest three-dimensional map of the 
universe over more than one-third of the sky. Luminous Red Galaxies (LRGs), 
Emission Line Galaxies (ELGs), and quasars (QSO) over the redshift range 
$0.05<z<2.1$, as well as the Lyman-$\alpha$ forest from QSO spectra from $2.1<z<3.5$,
will be used to trace the large-scale structure of the universe. 
DESI will target 48 million objects in order to provide key tests of cosmological 
models. The resulting map of the universe will chart the expansion history of the universe and the growth of structure through baryon acoustic oscillations (BAO; see e.g., \cite{Bassett2010} for a review) and redshift-space distortions (RSD; see e.g., \cite{Beutler2016, Grieb2016} for recent examples). 
Aside from dark energy, DESI will also open up broader investigations into 
cosmology and particle physics. For example, the scale dependence of the 
broadband power spectrum and the halo-mass dependent biasing will constrain 
primordial non-Gaussianity and thus inflationary models 
\cite{Gariazzo2015,Tellarini2016}. The damping of small-scale structure by
free-streaming neutrinos, affecting the broadband power spectrum, will enable precise
measurements of the sum of the neutrino masses \cite{Font-Ribera2014}.

To achieve these goals, DESI will look at 10000 (overlapping) patches of the 
sky, each 7.5 deg$^2$ in area. The total area covered by the survey will be 
14000 deg$^2$, and on average, a given point will be imaged approximately five 
times for a five-year survey. We define this number as the coverage, which can 
vary from zero to twelve at any particular spot for a five-year survey.
In the focal plane of the telescope, 5000 robotically-controlled fibers will 
observe the different objects visible in the tile among the 25000 or so in the 
focal plane, enabling massively parallel measurements of spectra and redshifts. The 
allocation of fibers to targets is optimized by a fiber-assignment algorithm that 
takes into account the necessary science requirements, such as the need to have 
multiple observations of quasars, while minimizing the number of unused fibers. 
However, the survey will still miss galaxies in very high-density regions, when 
all fibers have already been allocated. Furthermore, when only a single fiber 
covers two objects of different priorities, the lower-priority object will not 
be observed. As a consequence, only 78\% of ELGs are expected to be observed, 
compared to 95\% of LRGs and 98\% of QSOs (discussed in more detail 
in section \ref{sec:iso}). 

The inability to observe all targets can lead to systematic effects 
in the measured clustering signal. For example, a given point can only 
be observed on average five times, and in a high-density region such as 
a galaxy cluster, one expects several galaxies to remain unobserved. Since 
galaxy clusters have the highest bias, this leads to a real suppression of 
clustering. This is an effect that changes the 
apparent properties of observed objects with respect to the underlying targets 
but does not immediately signal a systematic error that needs to be 
corrected. It is best investigated using realistic simulations. 
However, incomplete target selection can also create signal or destroy 
existing signal. For example, if the coverage is spatially fluctuating, 
it leads to fluctuations in the angular structure even in the 
absence of any intrinsic clustering. Conversely, we will show 
that absorbing this effect by defining a spatially varying mean density 
typically leads to a suppression of existing angular clustering. 
We note that other observationally-based effects can also impact
the clustering signal, such as spectroscopic efficiency which is a 
function of fiber position in the focal plane and observation conditions, 
although such effects are not the focus of this work. 

Several complementary approaches for mitigating the effects of fiber 
assignment in past surveys have been developed. For example, the
clustering results from the Baryon Oscillation Spectroscopic Survey
(BOSS, \cite{Anderson2014,Anderson2014b,Alam:2016}) have demonstrated
the effectiveness of a nearest-neighbor weighting scheme, where the nearest
neighboring object on the sky is up-weighted to account for the
missing, fiber-collided objects. Alternatively, procedures to correct the 
effects of fiber collisions at the level of two-point clustering statistics have been
investigated for both the correlation function and the power spectrum 
\cite{Hawkins2003,Guo2012,Hahn2016}. 

In this article, we study the impact of the DESI fiber assignment on the measurement of the 
clustering of galaxies, investigating systematic errors in the galaxy 
power spectrum. We use DESI mock catalogs of ELG and LRG galaxies and 
quasars, as well as a simulated fiber assignment algorithm, and 
compute the power spectra of interest. We do not present methods to correct 
any fiber assignment artifacts, but rather examine the level of the
systematic errors introduced by fiber assignment and explore possible
estimators that can suppress these errors. We focus on ELGs in this work, 
as their observed fraction is the lowest. LRGs and QSOs have higher 
observability rates and one would expect the corresponding systematic effects 
to be smaller. However, this does not mean that these effects are negligible 
for LRGs and QSOs, and the lessons learned for ELGs cannot be immediately 
translated to LRGs or QSOs.

In a different, but complementary analysis, \cite{Burden2016} describes
a method to mitigate the effects of the DESI fiber assignment algorithm
on the correlation function of galaxies. While that analysis
focuses on clustering in configuration space, it finds similar effects 
on the clustering as we do for the power spectrum in Fourier space. 
It corrects for the effects of fiber assignment at level of the correlation 
function by presenting a modified clustering statistic, 
which relies on similar, but not identical, analysis methods as we discuss in
this work.

This paper is structured as follows. In section \ref{sec:the}, we define 
the fundamental physical quantities and give some useful properties of the 
power spectrum. In section \ref{sec:data}, we present the simulations. 
Section \ref{sec:survey} is dedicated to the survey design and the fiber 
assignment algorithm. The main results are presented in section \ref{sec:iso}, 
where we show the clustering results after fiber assignment for the power
spectrum multipoles (section~\ref{sec:aniso}) and $P(k,\mu)$ wedges 
(section~\ref{sec:pass5-wedges}), assuming a 5-year DESI survey.
We apply the same analysis to the first pass of DESI in section \ref{sec:pass1},
simulating the data available after only one year of the survey. Finally, we present 
some toy models of the fiber assignment on a periodic box for further analysis 
in section \ref{sec:box}, and conclude in section~\ref{sec:conclude}.

\section{Power spectrum in redshift space}
\label{sec:the}

DESI is a galaxy redshift survey and the basic ingredient of this study is the galaxy density field $n(\vec{r})$ which defines the overdensity field $\delta(\vec{r})$ and its Fourier transform $\tilde{\delta}(\vec{k})$:

\begin{equation}
\label{eqn:delta}
\delta(\vec{r})=\frac{n(\vec{r})-\bar{n}(\vec{r})}{\bar{n}(\vec{r})} ,
%\tilde{\delta}(\vec{k})&=\frac{(2\pi)^3}{V}\int %\delta(\vec{r})e^{i\vec{k}\cdot\vec{r}}d^3\vec{r}.
\end{equation}
where $\bar{n}(\vec{r})$ is the local mean density which, as we will show, needs to be carefully defined in a survey with varying coverage. We are interested in the two-point correlation of the overdensity, which is the 
correlation function $\xi(\vec{r})$ in configuration space and the power spectrum $P(\vec{k})$ in Fourier space. 

%\begin{subequations}
%\label{eqcorr}
%\begin{align}
%\xi(\vec{r})&=<\delta(\vec{r}+\vec{r'})\delta(\vec{r'})>_{\vec{r'}} \\
%P(\vec{k})&=\frac{(2\pi)^3}{V}<\tilde{\delta}(\vec{k})\tilde{\delta}^*%(\vec{k'})> \delta_D(\vec{k}+\vec{k'})
%\end{align}
%\end{subequations}

In this paper, we choose to work exclusively with the Fourier modes of the galaxy overdensity and 
the associated power spectrum, for reasons that will become clear later. 
Galaxies can be a biased tracer of dark matter, and in redshift space, 
we are sensitive to 
peculiar velocities of galaxies, so redshift-space distortions 
lead to observed anisotropic clustering. 
To describe the anisotropy, we define

\begin{equation}
\mu=\cos\theta=\frac{\vec{k}}{k} \cdot \hat{n},
\end{equation}
where $\theta$ is the angle between the wave-vector $\vec{k}$ and the line-of-sight direction $\hat{n}$. Rotational symmetry around the line-of-sight is assumed, and $P(k,-\mu)=P(k,\mu)$, so the anisotropic power spectrum 
can be expanded in even powers of $\mu$, or, in a Legendre series of 
even powers of $\ell$. For example, in linear theory the galaxy bias can be
expressed as \cite{Kaiser1987},

\begin{equation}
\tilde{\delta}(k,\mu) = (b_1 + f\mu^2)\tilde{\delta}^{r}_{m}(k)
\end{equation}
where $\tilde{\delta}^{r}_{m}(k)$ is the isotropic, matter overdensity field in real space, with corresponding power spectrum $P^r_m(k)$. The first term $b_1$ is the linear galaxy bias, and
the second term $f\mu^2$ accounts for RSD in linear perturbation theory, where $f=\mathrm{dln}D(a)/\mathrm{dln}a$ is the logarithmic derivative 
of the growth factor $D(a)$, where $a=1/(1+z)$. Hence, the power spectrum can be written as

\begin{equation}
P(k,\mu) = (1 + \beta \mu^2)^2 b_1^2 P^r_m(k),
\end{equation}
where $\beta = f/b_1$.

Non-linear effects, such as the so-called Finger-of-God effect, due to virialized velocities of galaxies in clusters (leading to elongated structures along the line-of-sight), can be modeled by multiplying the power spectrum by a Gaussian damping term $e^{-k^2\mu^2\sigma^2}$ \cite{Peacock1994, Park1994,Percival2009}.

The anisotropic power spectrum can be expanded into Legendre multipoles $\mathcal{P}_\ell(k)$, defined as

\begin{equation}
\label{eqn:multipoles}
\mathcal{P}_\ell(k) = \frac{2l+1}{2} \int_{-1}^{1} P(k,\mu)\mathcal{L}_\ell(\mu) d\mu.
\end{equation}
Here, $\mathcal{L}_\ell(\mu)$ is the $\ell^\mathrm{th}$ Legendre polynomial. In this basis, the monopole ($\ell=0$) represents the isotropic part of the power spectrum.
All multipoles with odd $\ell$ are equal to zero. We will be interested 
mostly in the low order multipoles (monopole, quadrupole, and
hexadecapole, or $\ell=0,2,4$), since measurement errors increase 
with increasing $\ell$, and there is little information left in
$\ell>4$ multipoles. However, we will 
describe below how this argument breaks down when dealing with 
systematics introduced by fiber assignment.

Conversely, one can compute $P(k,\mu)$ as a function of its multipoles. 
Using the previous properties, one can show that

\begin{equation}
\label{eqn:poles}
P(k,\mu)=\underset{\ell=0,2,4...}{\sum} \mathcal{L}_\ell(\mu)\mathcal{P}_\ell(k).
\end{equation}

If the sum is truncated after the third term, then one is only able to compute averages over a few, broad bins in $\mu$:

\begin{align}
	P \left[k,(\mumin,\mumax) \right]
   		&= \frac{\int_{\mumin}^{\mumax}P(k,\mu)d\mu}
        	{\int_{\mumin}^{\mumax}d\mu} \nonumber \\
     	&= \mathcal{P}_0(k) +\mathcal{P}_2(k) 
        	\times \frac{1}{2}\left[\frac{\mumax^3-\mumin^3}{\mumax-	
            	\mumin}-1\right] \nonumber \\ 
          &+ \mathcal{P}_4(k) \times \frac{1}{8}\left[\frac{7(\mumax^5-\mumin^5)-10(\mumax^3-\mumin^3)}{\mumax-\mumin}+3\right]. \label{eqn:pkmu-from-poles}
\end{align}
In this work, we will work mostly with three $\mu$ bins, equally spaced in $\mu$.

\section{Simulating the DESI observing strategy}
\label{sec:data}

We simulate the process of fiber assignment using realistic mock galaxy
catalogs for the DESI survey. Here, we describe the construction of those catalogs. 

\subsection{Simulations}

The mock catalogs are based on the \texttt{Outer Rim} N-body simulation \cite{Habib2012} within the HACC (Hybrid/Hardware Accelerated Cosmology Code) framework. %http://arxiv.org/pdf/1211.4864v1.pdf
The simulation employed $10240^3$ particles in a periodic box of 3 $\mathrm{Gpc}/h$ on 
a side with a flat $\Lambda$CDM model with $\Omega_m = 0.265$, $h = 0.71$,
and $\sigma_8=0.80$. 

The catalogs matching DESI survey volume contain $38$ million Emission-Line Galaxies, $4.7$ million Luminous Red Galaxies, and $2.7$ million Quasi-Stellar Objects. Each sample is generated by populating the dark matter halos from a single snapshot of the N-body simulation (see section \ref{sec:targets} for more details). This means that the clustering does not evolve with redshift. This assumption leads to a less realistic map of the sky, but it enables one to skip the complicated step of modeling the full light-cone.

In section \ref{sec:box}, we also use periodic box simulations, 
constructed using the Quick Particle Mesh (QPM) method \cite{White2014}, %http://arxiv.org/pdf/1309.5532v2.pdf
with a  flat $\Lambda$CDM model with $\sigma_8=0.826$. 
The resulting catalog corresponds to $143$ million galaxies in box of side length 5120 $\mathrm{Mpc}/h$. The box contains information about the positions in real space and the velocities of the galaxies. Adding velocities along a constant line-of-sight gives the corresponding box in redshift space, with perfectly plane-parallel RSD. This enables us to do our analysis in both real and redshift space, on a statistically significant sample, and with a well-defined line-of-sight across the periodic box.

%\begin{figure}[H]
%\begin{center}
%\caption{Geometry of the survey}
%\vspace{-0.3cm}
%\label{surveydesign}
%\end{center}
%\captionsetup{position=top}
%\subfloat[Tiling]{\label{tiling}\includegraphics[scale=0.32]{tiling.png}}
%\caption*{\footnotesize Tile centers for the DESI footprint, each pass %being represented with a different color. The spots do not indicate the %size of the area subtended by the focal plane. Figure taken from the DESI %Science Report.}\vspace{0.5cm}

%\subfloat[Fiber positioners]{\label{focalplane} %\includegraphics[scale=0.5]
%{focalplane.png}}
%\caption*{\footnotesize Fiber positioner locations for the full DESI %instrument. Fibers can move within the radius of one spot. Figure taken %from the DESI Science Report.}
%\end{figure}

\subsection{Survey Design and Fiber Assignment}
\label{sec:survey}

The design of the survey is optimized by the pattern of tiles on the sky, making the largest footprint outside the Milky Way that can be reached and observed well by the Mayall telescope. The tiling was obtained from a single covering of the whole sky by 28810 tiles, and then reduced to the actual accessible area. Eventually, 10666 tiles are distributed across 15789 deg$^2$, enough to encompass the final design of approximately 14000 deg$^2$. The footprint, composed of two separate regions, is covered by five passes of approximately 2130 tiles. 
%, as shown in figure  \ref{tiling}\,. 
Tiles often overlap and the mean coverage of a target is about five, with 
large fluctuations discussed further below.

The geometry of the survey is determined by this tiling and by the locations of the fiber positioners in the focal plane of the instrument. %shown in figure \ref{focalplane}\,. 
Each positioner can reach a galaxy within a patrol radius of $1.4'$, enabling a fiber assignment strategy based on science requirements. The combination of 5000 fibers  distributed in the focal plane of the telescope with each of the 10666 tiles gives rise to 54 million tile-fibers.
For further details regarding the final DESI instrument and science survey design, 
please see the Final DESI Design Report
\cite{DESIInstrumentDesign:2016,DESIScienceDesign:2016}.

\subsection{Targets}
\label{sec:targets}

DESI targets different types of objects. Because some types of objects require multiple observations, the fiber assignment algorithm establishes priorities for each of the targets, depending on its type. 

The lowest-redshift sample ($0.05<z<0.4$) of DESI is the Bright Galaxy Sample (BGS). Because it is surveyed during bright time when the sky is too luminous to allow observation of fainter targets, this sample does not fit in our study and BGS objects are not included in our mock catalogs.

Luminous Red Galaxies are massive red galaxies, which can be selected efficiently in a redshift range  $0.4<z<1.0$. The sample targeted by DESI is composed of 350 LRGs per square degree, 50 of which are expected to be misidentified, i.e., as stars or blue galaxies. Measuring their redshifts for $z>0.6$ can be challenging \cite{Levi2013}. They are the third highest-priority objects. In the simulation, the goal is to observe each LRG twice.

Emission-Line Galaxies are forming new stars at a high rate, with the massive young stars being responsible for their colors. With more than 2400 targets per square degree, they constitute the largest sample of DESI, over a wide range of redshift: $0.6<z<1.6$. Because of their high number density and relatively low bias, ELGs are chosen to be the lowest-priority objects and are observed only once.  The presence of the OII doublet at 373 nm facilitates their identification and redshift determination.

Quasars are used as density tracers for $z<2.1$ (QSO Tr) and also as backlights for clustering in the Lyman-$\alpha$ forest for $z>2.1$ (QSO Ly-$\alpha$). DESI will target 260 QSOs per square degree, 50 of which should be suitable for Ly-$\alpha$ forest measurements and 90 of which are expected to be falsely identified. Because of their point-like morphologies and their photometric characteristics close to the faint blue stars in optical wavelengths, measuring their redshifts is challenging. The Ly-$\alpha$ QSOs are given highest priority and we seek up to five observations of each. The QSO tracers are given second priority and we seek just one observation of each. 

The fiber assignment algorithm is also responsible for targeting Standard Stars (SS) for flux calibration and Sky Fibers (SF) for blank calibration. These use 2\% and 8\% of the fibers, respectively. 

\subsection{Algorithm}

For our study, we run a simulation of the fiber assignment algorithm. The DESI code that was used is publicly available at \url{https://github.com/desihub/fiberassign}.
It was run on the Cori computer at NERSC \cite{Cori} using a single node of 32 cores. The code is parallelized with OpenMP and typical run times are 30 minutes for assigning 48 million targets to 54 million tile-fiber combinations.

Below, we detail the different steps of the fiber assignment simulation algorithm.

%\newline\textbf{Read files:} \texttt{Targfile} lists the targets with their angular positions, their assumed types and corresponding properties (number of observations needed, priorities, etc.); \texttt{SStarsfile} displays the positions of the standard stars for calibration; \texttt{SkyFfile} shows the positions of the sky that can be used for blank calibration; \texttt{Secretfile} contains the precious informations about the true type and redshift of any target, it is read only for observed objects.

\begin{enumerate}

\item\textbf{Simple assign:}  Each fiber for each tile (that is, one of the 10666 pointings of the telescope) is associated with all the targets accessible to it using a kd-tree analysis. The priorities assigned to the various galaxy types are then used to make an initial assignment of each fiber.  The produces an overall collection of assignments where free fibers occur predominantly near the end of the survey.

\item \textbf{Redistribute tile-fibers, improve:} Tile-fibers are redistributed to other targets they can reach to spread the distribution of free fibers throughout the survey. Subsequent improvements increase the number of used fibers. Multiple iterations of
this redistribute/improve step optimize the allocation of tile-fibers to targets.

\item \textbf{Update types:} As the simulation proceeds, knowledge of the targets is updated according to previous observations: if the QSO target turns out to be a QSO tracer or not a QSO, then it needs no further observations; conversely, if it is a QSO for Ly-$\alpha$, then one wants to observe it up to four more times.

\item \textbf{Standard stars and sky fibers:} Remaining unused fibers are used for calibration if they are in an area where a standard star is available, or a sky measurement is possible.  Reassignments are made, if necessary, to make sure that the required number of standard stars and sky fibers are measured.
\end{enumerate}

\subsection{Outputs}

\begin{figure}[htbp]
\begin{center}
\includegraphics[scale=0.5]{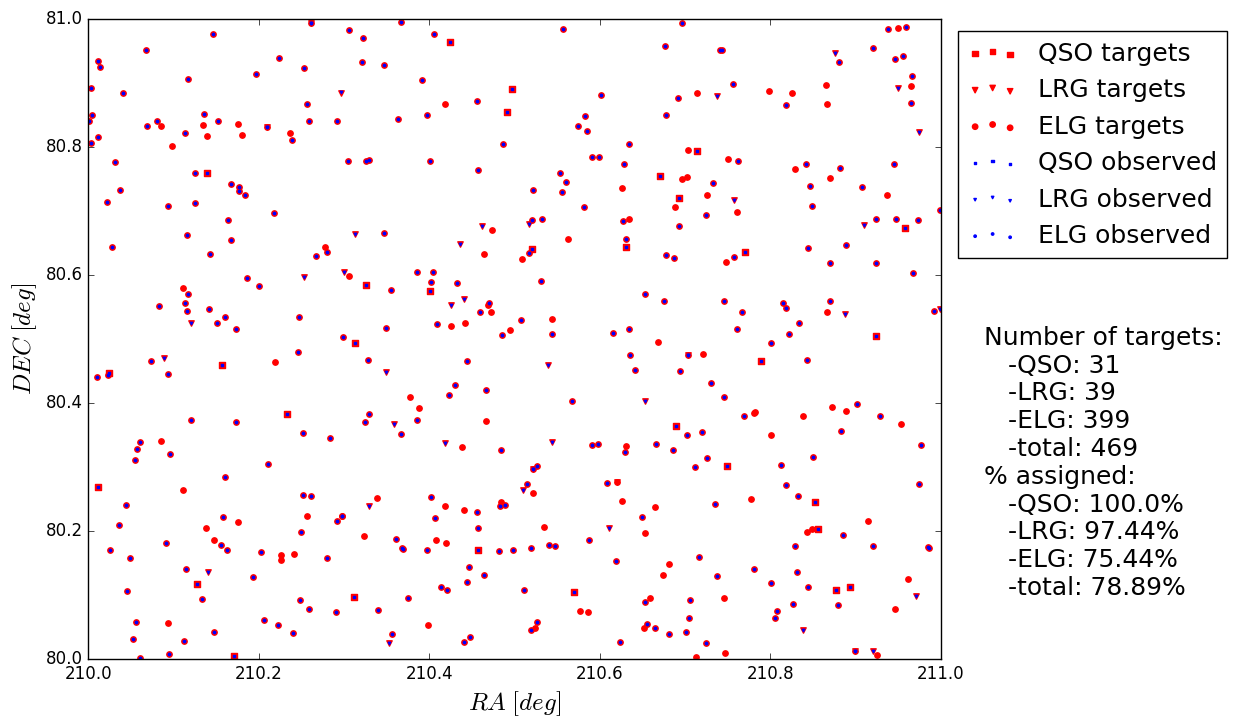}
\caption{A patch of one square degree of the sky. Shown are QSOs (squares), LRGs (triangles) and ELGs (circles). The targets are in red and if they are observed, they are overprinted in blue. The statistics of this particular patch are printed on the right.}\label{fig:mapsky}
\end{center}
\end{figure}

The output of the fiber assignment algorithm consists of 10666 files, one for each tile. Each file lists the objects observed by any of the assigned fibers, giving its type, sky coordinates, and redshift if it is a galaxy or a QSO. The algorithm also gives the relevant statistics per type of object, per square degree, such as the number of observations, percentage of rejection, etc. This is shown in table~\ref{tab:histofa}.  Figure \ref{fig:mapsky} represents a one square degree patch of the sky, with three kinds of targets. It shows which galaxies were observed, and the statistics over this particular patch. In total, 94\% of the fibers were assigned, so each tile observes on average 4700 objects. While the ELG sample is the largest, they also have the lowest-priority; as a result, out of 38 million targets, only 29 million are assigned to a fiber and have their redshifts measured.

The most notable features of the fiber assignment algorithm are that the assignment 
misses more galaxies in overdense regions and that the selection is 
done in the plane perpendicular to the line-of-sight. These features introduce 
a modulation of the density field, and the main goal of our study is understanding this modulation.

\begin{table}[tb]
\centering
\small
\setlength\tabcolsep{2pt}
\begin{tabular}{|c|c|c|c|c|c|c|c|c|c|c|}
\hline
Type & 0 obs. &   1  obs.&   2 obs. &  3 obs.&  4 obs.&  5 obs. &initial&  fibers&obs.& $\%$obs.\\ \hline
    QSO (Ly-$\alpha$)   &   0 &     2 &   7 & 15 & 20 & 3 &    49 &   161 &    49 & 98 \\ \hline
  QSO (Tr)   &   1 &   112 &   4 &  0 &  0 & 0 & 119 &   123 &   117 & 98 \\ \hline
        LRG   &  15 &    47 & 235 &  0 &  0 & 0 & 298 &   518 &   283 & 95 \\ \hline
        ELG   & 525 & 1885 &   0 &  0 &  0 & 0 & 2411 & 1885 & 1885 & 78 \\ \hline
    Fake QSO   &   1 &    84 &   3 &  0 &  0 & 0 & 90 &    92 &    88 & 98 \\ \hline
    Fake LRG   &   2 &    44 &   2 &  0 &  0 & 0 & 50 &    50 &    47 & 95  \\ \hline
\end{tabular}
\caption{The statistics of the fiber assignment algorithm per square degree. The first five columns show how many objects were observed a certain number of times, for each type. Then, there are the numbers of targets, assigned fibers, observed galaxies and percentage of observed objects, again for each type.} \label{tab:histofa}
\end{table}

\section{Power spectrum measurements} \label{sec:iso}

In this section, we present results of the power spectrum analysis on the
DESI geometry. We begin with the multipole 
analysis, followed by the projection of multipoles into $P(k,\mu)$ wedges 
(bins in $\mu$).

Before presenting the results we need to discuss the weighting of the 
galaxies. The local mean density, $\bar{n}(\vec{r})$, 
is needed to compute the physical overdensity field (equation \eqref{eqn:delta}). 
The simplest weighting one can employ is inverse  noise weighting. Noise
in a galaxy survey is determined by the shot noise power spectrum 
$N$, which in the simple Poisson model, is given by the inverse of the local mean number density $N=\bar{n}(\vec{r})^{-1}$. 
In this case, the inverse noise weighting gives $\delta(\vec{r})/N=n(\vec{r})-\bar{n}(\vec{r})$. 
The Fourier mode is given by $\delta(\vec{k}) =\int [n(\vec{r})-\bar{n}(\vec{r})] \exp(i\vec{k}\cdot \vec{r})dV$. 
Note that the integrals become simply a sum over all 
observed galaxies minus the mean number of galaxies. 
The latter is usually written in terms of the number of random 
galaxies that have been generated in the absence of any clustering, 
where the number of randoms can be over-sampled to reduce their 
shot noise. Hence, the expression involves a sum over all observed 
galaxies minus the sum over all the random galaxies, multiplied by 
the Fourier term $\exp(i\vec{k} \cdot \vec{r})$. This is the 
Fourier space version of the Landy-Szalay estimator for the correlation function \cite{Landy1993}. Since 
we have inverse noise weighting the expression is not yet 
properly normalized, so in the end, one must also divide by the 
sum of the weights squared, multiplied by the volume, which gives the appropriate normalization factor \cite{FKP1994}. 

Inverse noise weighting is valid for estimating the power on small 
scales, where the clustering power is small. On large scales, where 
the clustering power is large compared to the noise, uniform weighting is a 
better choice. A more general weighting scheme is the so-called FKP weighting \cite{FKP1994}, where 
the inverse noise weighting is given by the sum of signal and noise 
$N+P(k_0)$, where $P(k_0)$ is the fiducial power at a 
scale at which one is estimating the power spectrum. In this 
case, the sum over all galaxies and randoms is multiplied by 
$w_{\rm FKP}=[1+\bar{n}(\vec{r})P(k_0)]^{-1}$. To evaluate this one 
needs an estimate of $\bar{n}(\vec{r})$. 

In this paper, we will only explore pure inverse noise weightings, 
and we will not investigate the more general FKP weighting. Adding
FKP weighting should not pose any problems for the methods where 
the local mean is determined, which is our method of choice. It 
can be more of an issue for some of the simpler methods we also 
discuss in this paper. 

To compute the multipoles of the power spectrum, we used the publicly-available 
\texttt{nbodykit}\footnote{https://github.com/bccp/nbodykit} 
software package \cite{nbodykit}, which contains 
an implementation of the power spectrum estimator described in \cite{Bianchi2015,Scoccimarro2015}.
The algorithm first transforms coordinates from a spherical system of angles and redshifts into a cartesian comoving system $\vec{r}$ assuming a fiducial cosmology. It then computes the density field $n(\vec{r})$.
We also need the mean number of galaxies per unit volume at a given redshift $\bar{n}(\vec{r})$, which 
can be determined by taking the total 
number of galaxies per unit redshift $\mathrm{d}N/\mathrm{d}z$, 
and dividing by the 
volume of the survey per same unit redshift $f_\mathrm{sky} dV/dz$. Here,
$dV/dz=4\pi \chi^2 d \chi/dz$ (flat cosmology assumed) 
can be determined from the comoving distance -- redshift relation 
assuming a fiducial cosmology, 
and $f_\mathrm{sky}$ is the area of the survey as a fraction of 
the whole sky area. Typically, 
one smooths the observed $dN/dz$
to suppress radial fluctuations generated by the large-scale structure. 
The area of the survey is defined as the area where the probability of observing 
a galaxy is above zero. 

The procedure above assumes the mean density does not vary 
in the angular direction. As discussed in the previous 
section, this assumption does not hold for DESI, as the 
coverage can vary significantly across the sky, 
as a consequence of the fiber assignment procedure. 
This effect leads directly to non-cosmological fluctuations in the observed number density of objects and must be accounted for.

\subsection{Defining local mean density with randoms}\label{sec:randoms-defs}

As described above we need to define the local mean density
$\bar{n}(\vec{r})$. 
The basic idea is to use a Monte-Carlo simulation by placing random galaxies in the DESI survey according to the properties one wants to give them. We must also normalize the number 
of randoms to the total number of observed galaxies (the so-called integral constraint). In the standard analysis, the total number of randoms 
can and should be significantly larger than the number of observed galaxies (so that the shot noise from randoms is suppressed), and each is given a weight $w=\Ngal/\Nran$,  
so 
that the sum over $\Nran$ randoms equals the total number of observed 
galaxies $\Ngal$. We now 
define several different randoms that we can use in the 
context of DESI. 

\subsubsection{``Uniform" Randoms, $R_u$}\label{sec:uniform-randoms}

The simplest random catalog one can define simply 
uses objects whose angular positions are distributed randomly within the
survey geometry and whose redshifts are drawn randomly from the
same distribution as the observed galaxies. We will denote this case as \texttt{randoms\_uniform}, or $R_u$. Their weight is given 
by $w=\Ngal/\Nran$. 
This type of randoms reproduces the redshift selection of the galaxies. 
They are uniform in the angular direction, and hence do not 
quantify the variable coverage created by fiber assignment. Since fiber assignment 
can imprint structure on very large scales, it 
can create spurious correlations in the angular direction. 
Although we can use this randoms catalog to compute the clustering of the 
targets catalog, we will show that these randoms fail to recover the 
true monopole of the density field after fiber assignment. 
%This is shown in Figure \ref{fig:iso}. As expected, these randoms produce a huge excess of power at low $k$. 

\subsubsection{``Weighted" Randoms, $R_w$}\label{sec:weighted-randoms}

To remedy the issues with \texttt{randoms\_uniform}, 
we introduce a new version of randoms, where we 
normalize them separately for each coverage. We denote this case as 
\texttt{randoms\_weighted}, or $R_w$.
To be more specific, we define the local coverage $i$ of a galaxy (data or random) to be the number of fibers from different tiles that can reach it. This definition does not mean that if the coverage is greater than one, the object is observed more than once, but that it could have been visited $i$ times. For example, an ELG can be covered up to 12 times, but can only be observed once, or not at all. Then, we denote $\Ngal(i)$ to be the total number of \emph{observed} true galaxies with a coverage $i$, and $\Nran(i)$ to be the total number of \emph{}{targeted} random galaxies with a coverage $i$. The individual weight of a random covered $i$ times is defined as $w_i=\Ngal(i)/\Nran(i)$. 
With these weights, \texttt{randoms\_weighted} defines properly the mean number of galaxies in a location of coverage $i$, as a function of $i$, and thus takes into account the effects of the fiber assignment as a function of coverage on the local mean number density. The statistics of sky coverage and corresponding weights are shown in table~\ref{tab:weights}. 
Here, we have started with the 
number of randoms equal to the number of target galaxies, so for 
a high coverage number 
$i$, we expect the weights to be close to unity, since most of the targets 
will be observed, while for low 
coverage $i$ we expect it to fall significantly below unity, 
because only a small fraction of targets will actually be observed. 

\subsubsection{Other Randoms, $R_{fa}$ and $R_s$}\label{sec:other-randoms}

We have also explored two other types of simplified random catalogs, 
which we show below to be inferior to \texttt{randoms\_weighted}, 
but which can sometimes be simpler to evaluate. The first is denoted as 
\texttt{randoms\_switch} or $R_s$. In this case, the objects
have the same angular position as the observed galaxies but random redshifts drawn from the same distribution as the data. Thus, their angular correlation is the same as the galaxies (reproducing perfectly the effects of the fiber assignment), but their three-dimensional correlation should be randomized. Using them introduces a perfect correlation between data 
and randoms in the transverse plane perpendicular to the line-of-sight, effectively
removing these angular modes from the clustering measurement. 
They can be over-sampled in the radial direction, reducing shot noise, in which 
case they also can be weighted by $w=\Ngal/\Nran$. This type of randoms catalog
is used extensively in the work presented in \cite{Burden2016}.

A second set of simplified randoms we explore is called
\texttt{randoms\_after\_fa}, or $R_{fa}$. They are constructed 
by passing a \texttt{randoms\_uniform} catalog through the fiber
assignment algorithm. It is important that the initial
number of randoms is the same as the number of real targets. 
Because of the non-linearity of the fiber assignment selection, fields with different
levels of fluctuation are affected differently by the algorithm. Thus, the
recovered mean density after applying fiber assignment to uniform randoms differs systematically from the true mean, as we will show in the following sections. 
The rate of rejection for randoms as an input of the fiber assignment is somewhat smaller than the one for true galaxies due to the absence of clustering. To reduce 
shot noise in the randoms catalog, one can repeat the fiber assignment 
procedure on several copies of randoms and then average them together. 
The results presented here only use a single copy of randoms, although tests with 
multiple copies yield similar trends and results. In the end, the objects
in this type of randoms catalog also receive a weight $w=\Ngal/\Nran$. 

%A summary of the different kinds of randoms and their main characteristics are presented in table \ref{tablerandoms}\,, together with 

%\begin{table}[H]
%\centering
%\caption{Different kinds of randoms}
%\label{tablerandoms}
%\begin{tabular}{|c|c|c|c|c|}
%\hline
%Symbol & Name      & Angular position         & Redshift                 %distribution  & Weight  \\ \hline
%$R_u$ & \texttt{randoms\_uniform}   & Random                   & %$N_{gal}%(z)$ &1 \\ \hline
%$R_s$ & \texttt{randoms\_switch}    & Same as galaxies after FA       & %$N_{gal}(z)$ & 1 \\ \hline
%$R_{fa}$ & \texttt{randoms\_after\_fa} & Random passed through FA & %$N_{gal}(z)$ & 1 \\ \hline
%$R_w$ & \texttt{randoms\_weighted} & Random & $N_{gal}(z)$ &  $N_{gal}(i)/N_{ran}(i)$ \\ \hline
%\end{tabular}
%\vspace{0.3cm}
%\caption*{\footnotesize The four different kinds of randoms and their %important characteristics}
%\end{table}

\begin{table}[bt]
\centering
\footnotesize
\setlength\tabcolsep{2pt}
\begin{tabular}{|c|c|c|c|c|c|c|c|c|c|c|c|c|c|}
\hline
$i$ & 0 & 1& 2& 3& 4& 5 &6& 7& 8& 9& 10& 11& 12 \\ \hline
$\Ngal(i)$& 0 &95826& 248175 &939743 &4571219 &12757734 &8973518& 1860511 &305687& 30916& 3252& 398 &21 \\ \hline
$\Nran(i)$ &70268& 643141& 769134& 1761988& 6589071& 15975651 & 9982090& 1937284 &309683& 30982 &3254 &425 &9  \\ \hline
$w_{i}$ & 0 &0.149 &0.323 &0.533 &0.694& 0.799 &0.900& 0.960& 0.987& 0.998& 0.999& 0.936& 2.111 \\ \hline 
\end{tabular}
\caption{Coverage $i$ and the corresponding weights $w_i$ for \texttt{randoms\_weighted}. Total number of objects is $38 \times 10^6$. Note that 0.18\% of the survey area is not accessible at all (0 coverage).} \label{tab:weights}
\end{table}

\begin{figure}[hbtp]
	\centering
    \includegraphics[width=0.8\textwidth]{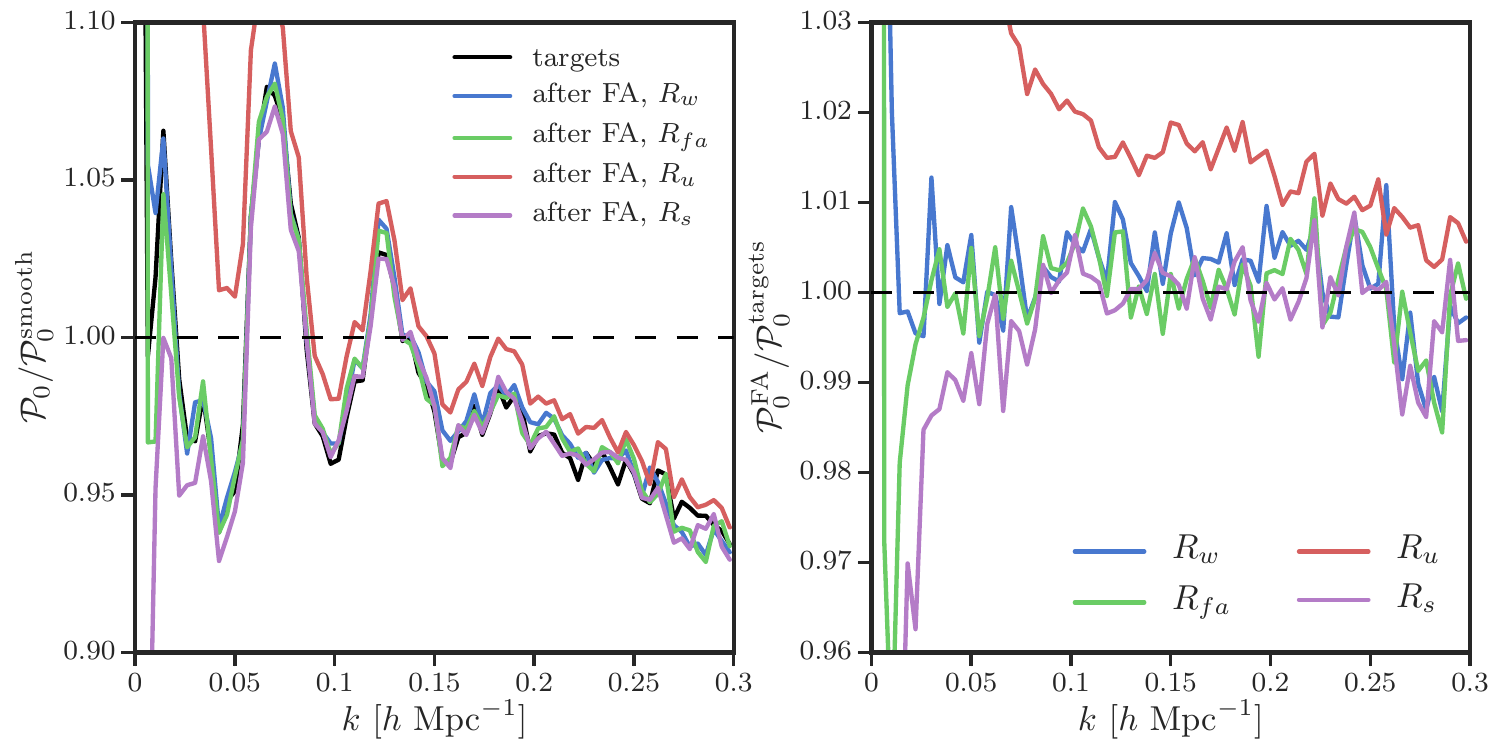}
  \caption{Left: the target monopole power spectrum $P_0(k)$ and the corresponding 
  monopole power spectra after fiber assignment using different randoms (as described
  in section~\ref{sec:randoms-defs}), divided by a smooth no-wiggle power spectrum. 
  Right: the ratios of $P_0(k)$ after fiber assignment using the different randoms, 
  divided by the target $P_0(k)$. Results are shown for the full 5-year DESI survey.}
\label{fig:iso}
\end{figure}

% \begin{figure}[hbtp]
%   \begin{subfigure}[b]{0.5\textwidth}
%     \includegraphics[width=\textwidth]{ELG_mono_wiggles_compare_4_randoms.png}
%     \caption{}\label{fig:elgmonowiggles}
%   \end{subfigure}
%   %
%   \begin{subfigure}[b]{0.5\textwidth}
%     \includegraphics[width=\textwidth]{ELG_ratios_mono_compare_4_randoms.png}
%     \caption{}\label{fig:elgmonoratios}
%   \end{subfigure}
%   \caption{(a) target monopole power spectrum $P_0(k)$ and the corresponding monopole power spectrum after fiber assignment using different randoms, divided by a smooth no-wiggle power spectrum. (b) ratios of $P_0(k)$
% after fiber assignment using different randoms, divided by the target $P_0(k)$. We also show the mean of the ratio, averaged over all $k$ shown.} \label{fig:iso}
% \end{figure}

\subsection{Multipoles}
\label{sec:aniso}

In figure \ref{fig:iso}, we compare the ability of the four different kinds of random catalogs to recover the monopole $\ell=0$ of the target catalog with only galaxies passed through fiber assignment. We see that \texttt{randoms\_uniform} fails to reproduce the local mean number density because even the isotropic power spectrum is incorrect. Using the three other kinds of randoms enables precise isotropic measurements, with an error $<1\%$ as shown in figure \ref{fig:iso}, except at very low $k$, 
where \texttt{randoms\_weighted} shows the least deviation
from the true power spectrum.

We proceed with the results for higher order multipoles
$\mathcal{P}_\ell$, for $\ell=2,4$. 
Computing these multipoles for a wide-angle galaxy redshift survey is challenging because the line-of-sight direction 
varies across the survey, and we cannot assume a plane-parallel 
approximation. We use the FFT-based algorithm of \cite{Bianchi2015,Scoccimarro2015} to compute these higher-order multipoles, with the quadrupole and hexadecapole requiring four and sixteen FFTs, respectively.

These higher-order multipoles enable us 
to describe the anisotropic power spectrum and determine the effects of redshift-space 
distortions. By comparing the results before and after fiber assignment, we can quantify
the level of anisotropy artificially introduced by fiber assignment and thus the level of systematic bias.
  
\begin{figure}[t]
\centering
    \includegraphics[width=0.8\textwidth]{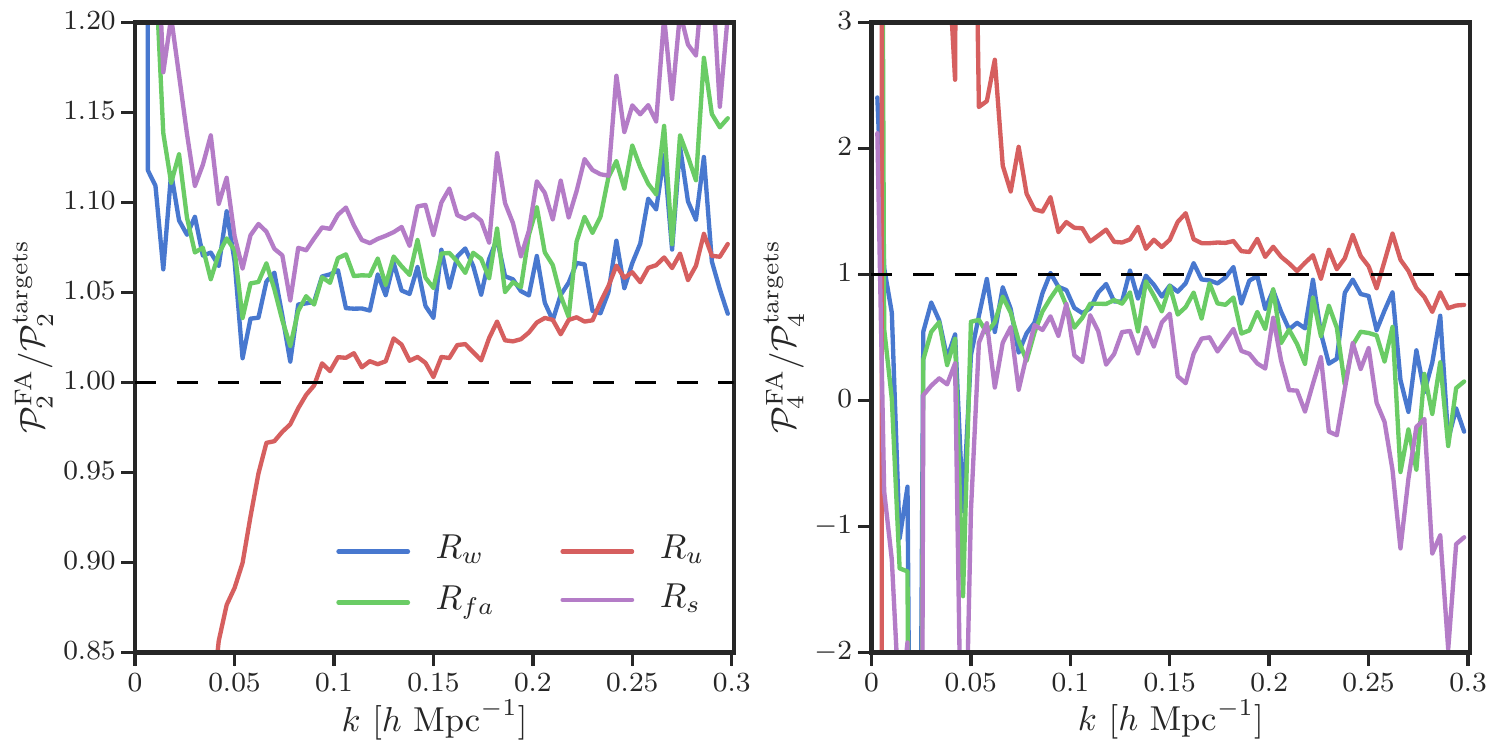}
  \caption{The ratios of the quadrupole (left) and hexadecapole (right) after 
  fiber assignment to the same quantities measured from the targets catalog for the
  full 5-year DESI survey. We compare the ratios for each of the definitions of 
  randoms described in section~\ref{sec:randoms-defs}.}\label{fig:aniso1}
\end{figure}

In figure \ref{fig:aniso1}, we compare the effects of fiber assignment, using different random catalogs, on the quadrupole and hexadecapole. Once 
again,
\texttt{randoms\_weighted} have the least deviation from the 
true power spectrum and hence is the best solution to recover the first three multipoles of the power spectrum. However, the anisotropies induced by fiber assignment are always important, and randoms alone 
cannot remove the problem. Indeed, we have an error of approximately $6\%$ on the quadrupole (averaged over all $k$) and approximately $40\%$ on the hexadecapole, which obviously cannot be neglected. We note that \texttt{randoms\_after\_fa} results are not significantly worse. 

\subsection{Power spectrum binned in $\mu$}\label{sec:pass5-wedges}

The multipoles are not necessarily the best basis to describe the anisotropies of the power spectrum induced by fiber assignment. Because target 
selection is done in the purely angular direction, we expect most of the 
effects of fiber assignment to be localized to transverse modes, with $\mu=0$. In this 
section, we discuss results from a binned $\mu$ analysis. 

\begin{figure}[btp]
\centering
    \includegraphics[scale=0.8]{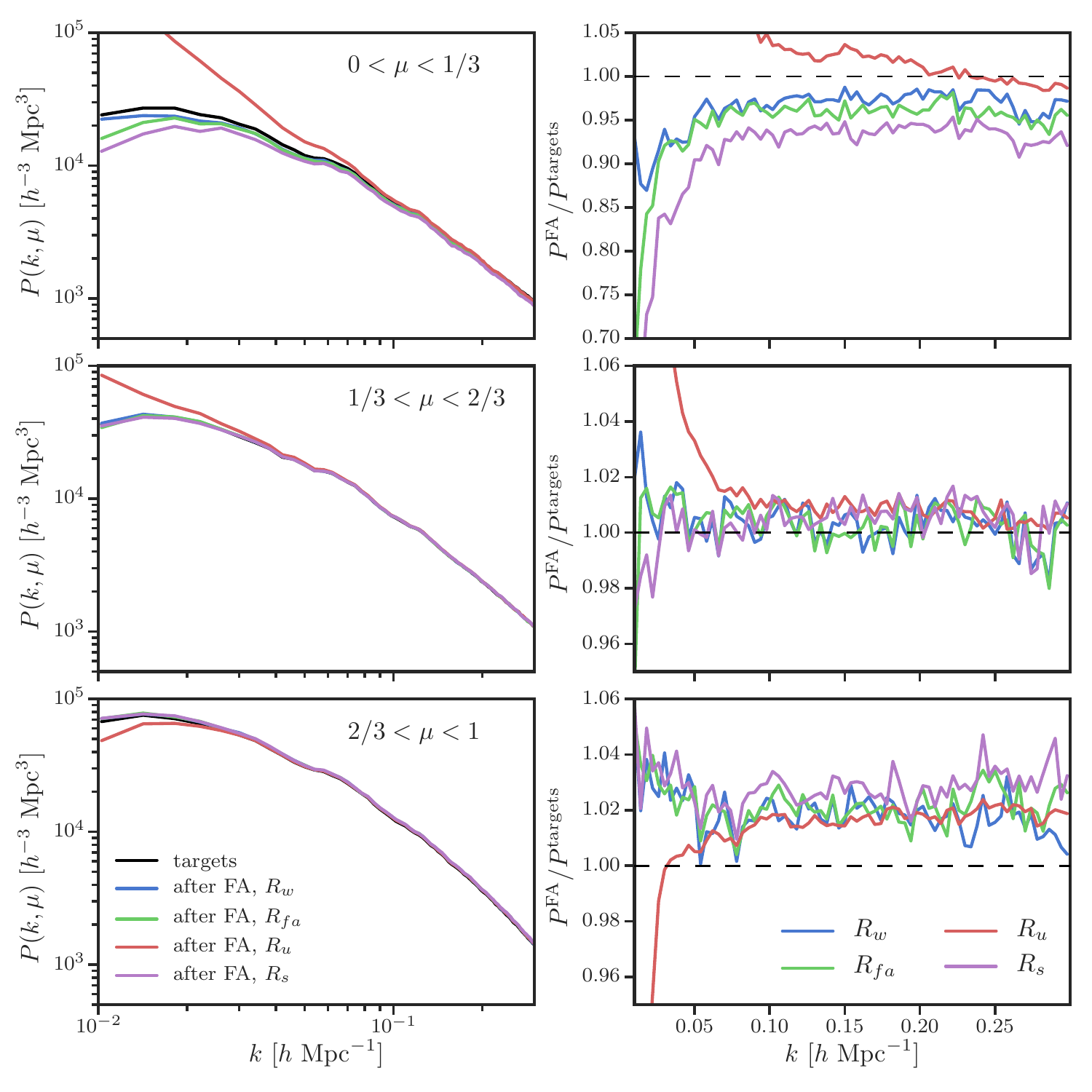}
\caption{The measured results for 3 $P(k,\mu)$ wedges after fiber assignment, 
		as computed from the $\ell=0,2,4$ multipoles, for the full DESI survey. 
        We compare the ratios (right) to the target wedges for each of the 
        definitions of randoms described in section~\ref{sec:randoms-defs}.
        We find the best results using the coverage-weighted randoms, $R_w$.}
\label{fig:pkmu}
\end{figure}

As discussed in section \ref{sec:the}, the anisotropy of the power spectrum can be described by the cosine of the 
angle between the Fourier mode 
and the line-of-sight, $\mu$. Using the multipoles of targets and galaxies after fiber assignment with different randoms, as presented in the previous subsection, we compute $P(k,\mu)$ for three $\mu$ bins according to equation \eqref{eqn:poles}. The results are presented in figure \ref{fig:pkmu}. Note that the binned values are not exact since we assume only the first three multipoles are non-zero, 
an assumption that would be explicitly violated if we have a 
significant $\mu=0$ term in addition to the three multipoles that 
arise from linear theory. Still, we expect that such analysis 
would reveal the existence of such a component. 

As shown in figure \ref{fig:pkmu}, the \texttt{randoms\_weighted} 
catalog gives the best results. 
Most of the effects of the fiber assignment on large scales are in the 
lowest $\mu$ bin, i.e. $k_\parallel\sim 0$ modes.
This can be understood by the fact that fiber assignment selects targets in the plane perpendicular to the line-of-sight, hence affecting the $\mu=0$ modes. We see that uniform randoms give excess power, while the other 
randoms suppress power in this bin. As discussed above, the 
spatially varying coverage factor creates clustering not accounted for
by uniform randoms. For the other three randoms just the opposite happens: 
in the limit where the coverage is a lot smaller than the average 
number of targets using the local mean
completely destroys clustering in the 
transverse direction. 

For the two other bins the errors are very small: sub-percent error for the middle $\mu$ bin 
and a $2\%$ enhancement 
for the highest $\mu$ bin. One expects fiber assignment to remove 
high density peaks, since the maximum number of objects 
that can be observed within a patrol radius is given by the mean coverage 
(with an average of five), while in a dense region like a cluster 
one expects to have more targets. This would reduce the cluster 
contribution, and since clusters are more strongly biased, it 
would reduce the clustering amplitude. We do not observe this effect, 
but it is not clear whether this is due to the limitations of our 
analysis (such as using multipoles up to $\ell=4$ only), or 
due to the fact that ELGs have a broad redshift distribution and are not strongly 
clustered, reducing this effect. 

These plots give an idea of how the fiber assignment affects the anisotropy of 
galaxy clustering. Evidently, we can restrict the contamination
to the lowest $\mu$ bin when using the appropriate randoms. 
However, one would also like to compute $P(k,\mu)$ directly in 
narrow $\mu$ bins to confirm this result. The power spectrum estimator
used in this case (in the presence of the survey geometry) can 
only measure the $\ell=0,2,4$ multipoles, and hence we turn to alternative methods. 

\subsection{$P(k,\mu)$ on a small box}

One possible alternative is to perform 
an FFT analysis on a small cubic box, which enables us to compute directly $P(k,\mu)$ in Cartesian coordinates (X,Y,Z). We are faced with two issues: the DESI survey has a complex geometry very different from that of a periodic box, and the local mean number density varies significantly across the survey (hence the crucial role of randoms). Such analysis is usually done on a periodic box with a constant mean number density. In that case, no randoms are needed because $\bar{n}(\vec{r})=\bar{n}=N_\mathrm{tot}/V_\mathrm{tot}$.

First, we pick a small box in the middle of the DESI survey such that it is completely filled, being careful with its orientation such that we first choose the $Z$ direction of the box to define its line-of-sight, and orient the box so that the $X-Y$ plane is perpendicular to the chosen line-of-sight. The line-of-sight is assumed to be constant along the box, which is true only if the box is sufficiently small and far from the observer, i.e. seen through a small solid angle. The box we use here is a cube of length 1 $\mathrm{Gpc}/h$, centered on $z=1$, with the $Z$ axis as the line-of-sight. The box contains 0.55 million targets, representing only 1.5\% of the total survey. However, one should be aware that the plane-parallel approximation is not completely valid, since the box covers $(27 \deg)^2$ on the sky.

\begin{figure}[btp]
\centering
	\includegraphics[width=0.7\textwidth]{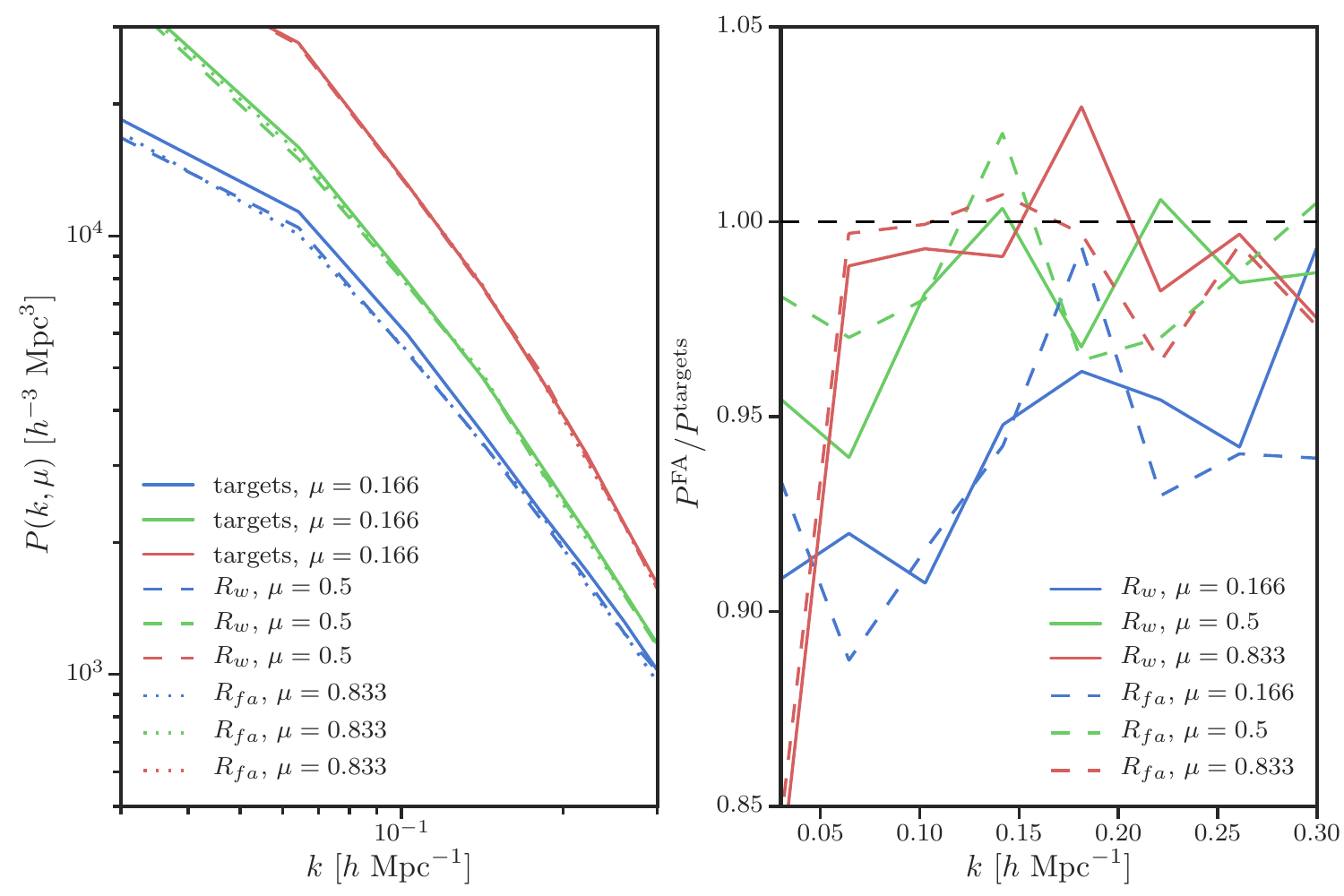}
\caption{$P(k,\mu)$ results for three $\mu$ bins on a small box cut out of the full
DESI volume. We show the unnormalized spectra (left) as well as the ratio of
the power after fiber assignment to the targets spectra (right). As there are fewer 
modes than in the full survey analysis, the ratios are noisier.}
\label{fig:smallboxpkmu}
\end{figure}

We developed a new algorithm, similar to the standard periodic box FFT power spectrum analysis, which can also take randoms as an input to compute the local mean and the corresponding overdensity field as described previously. With this, we should be able in principle to compute $P(k,\mu)$ for an arbitrary number of $\mu$ bins. However, the 
variation of the line-of-sight direction across the box still yields an error on the 
measured $\mu$ values, which then no longer preserves the purity of 
transverse modes. 
Therefore, we perform the analysis for only three bins (figure \ref{fig:smallboxpkmu}). We only use \texttt{randoms\_weighted} and 
\texttt{randoms\_after\_fa}, since they performed considerably better 
than the alternatives in the tests above. The effects of fiber assignment seem to be again localized to the lowest $\mu$ bin. The change of power for the two
higher bins is less than 2\%. This supports the idea that fiber assignment causes the loss of power in the plane perpendicular to the line-of-sight, and that all other directions are minimally affected.

\subsection{First year of DESI}
\label{sec:pass1}

 The goal of this section is to simulate the data analysis that could be done with data collected in the first year of the DESI survey. We assume that the first year covers the entire footprint once. We call this type of coverage ``pass 1''. With the circular shape of the focal plane, inevitably some spots will not be covered and others will have received multiple coverage.  We utilize the fiber assignment algorithm with the priorities among LRG, ELG, and QSO as described above.  The methods are exactly the same as the ones presented in previous sections, so we present only the main results.

Running the fiber assignment simulation for pass 1 instead of the full survey is relatively easy given the pipeline of the code. Indeed, we only need to run the assignment for $2139$ tiles corresponding to pass 1, out of the $10666$ total. The first pass is very unfavorable for ELGs. Indeed, as they are the lowest-priority objects, most of them are observed later in the survey, once other objects have already been observed, sometimes several times.  In the output of the fiber assignment, we read that 11.6\% of ELGs were observed after pass 1, against 78.2\% for the full survey. The overall statistics of the fiber assignment for pass 1 only are shown in table~\ref{tab:stats2}. Measuring the true power spectrum with these data is thus much more challenging. However, we show in the next paragraph that our analysis already enables reasonably good power spectrum measurements.

\begin{table}[b]
\small
\centering
\begin{tabular}{|c|c|c|c|c|c|c|c|c|c|c|}
\hline
Type & 0 obs. &   1  obs.&   2 obs. &  3 obs.&  4 obs.&  5 obs. &initial&  fibers&obs.& $\%$obs.\\ \hline
QSO (Ly-$\alpha$)   &    22 &  25 &  1 & 0 & 0 & 0 &    49 &  29 &  27 & 55.2\\ \hline
QSO (Tr)   &    53 &  66 &  0 & 0 & 0 & 0 &   119 &  66 &  66 & 55.1 \\ \hline
LRG   &   151 & 137 & 10 & 0 & 0 & 0 &   298 & 157 & 147 & 49.5 \\ \hline
ELG   & 2,130 & 280 &  0 & 0 & 0 & 0 & 2,411 & 280 & 280 & 11.6\\ \hline
Fake QSO   &    40 &  49 &  0 & 0 & 0 & 0 &    90 &  49 &  49 & 55.1 \\ \hline
Fake LRG   &    24 &  25 &  0 & 0 & 0 & 0 &    50 &  25 &  25 & 51.0  \\ \hline
\end{tabular}
\caption{Statistics of the Fiber Assignment for pass 1 only, in $1$deg$^2$.}\label{tab:stats2}
\end{table}

\begin{figure}[tb]
\centering
    \includegraphics[width=0.8\textwidth]{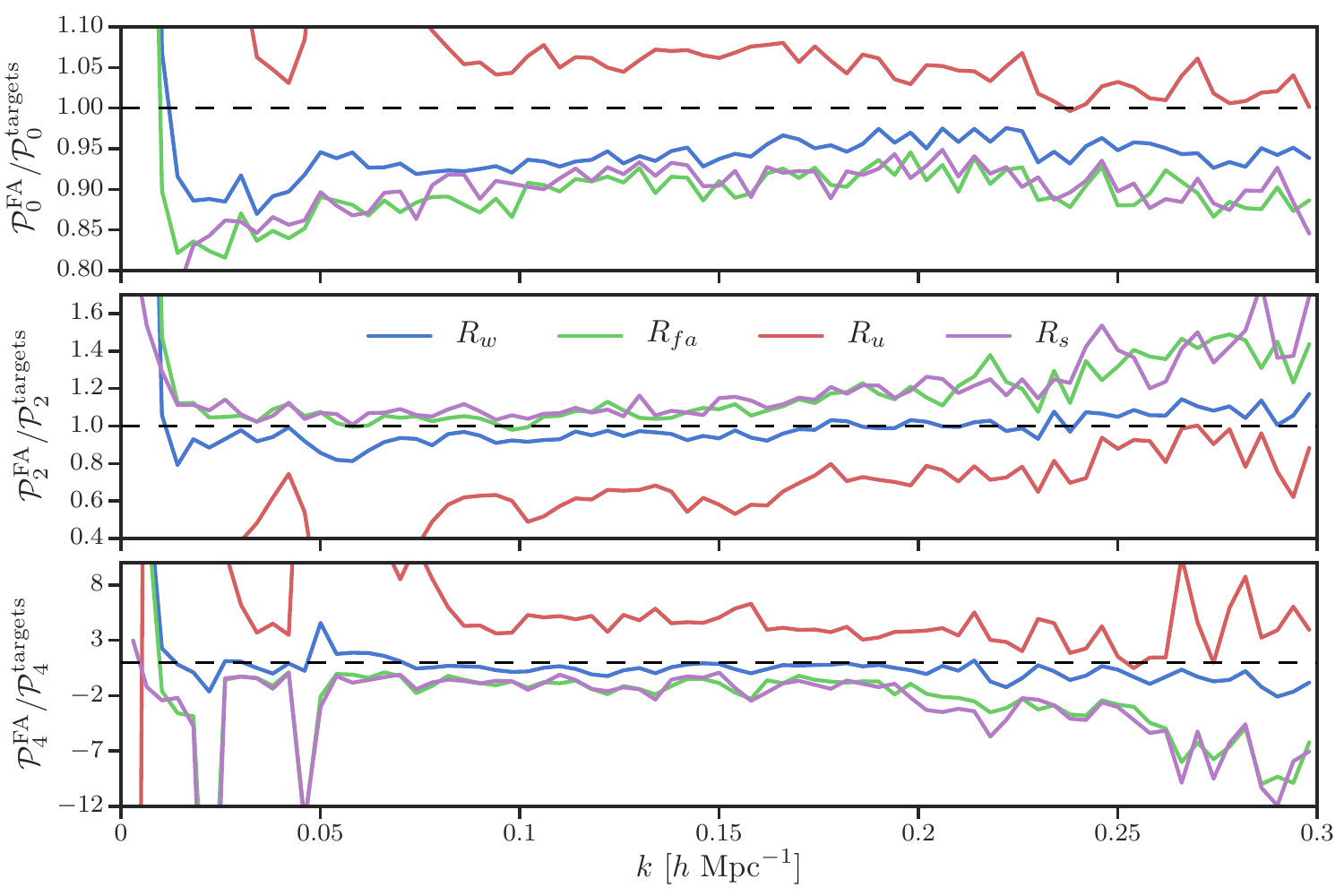}
\caption{DESI pass 1 results: the ratios of the monopole (top), quadrupole (middle), and 
	hexadecapole (bottom) after fiber assignment to the same quantities measured 
    from the targets catalog. We compare the 
    ratios for each of the definitions of randoms described in 
    section~\ref{sec:randoms-defs}.}\label{fig:pass1poles}
\end{figure}

\begin{figure}[bt]
\centering
    \includegraphics[width=0.8\textwidth]{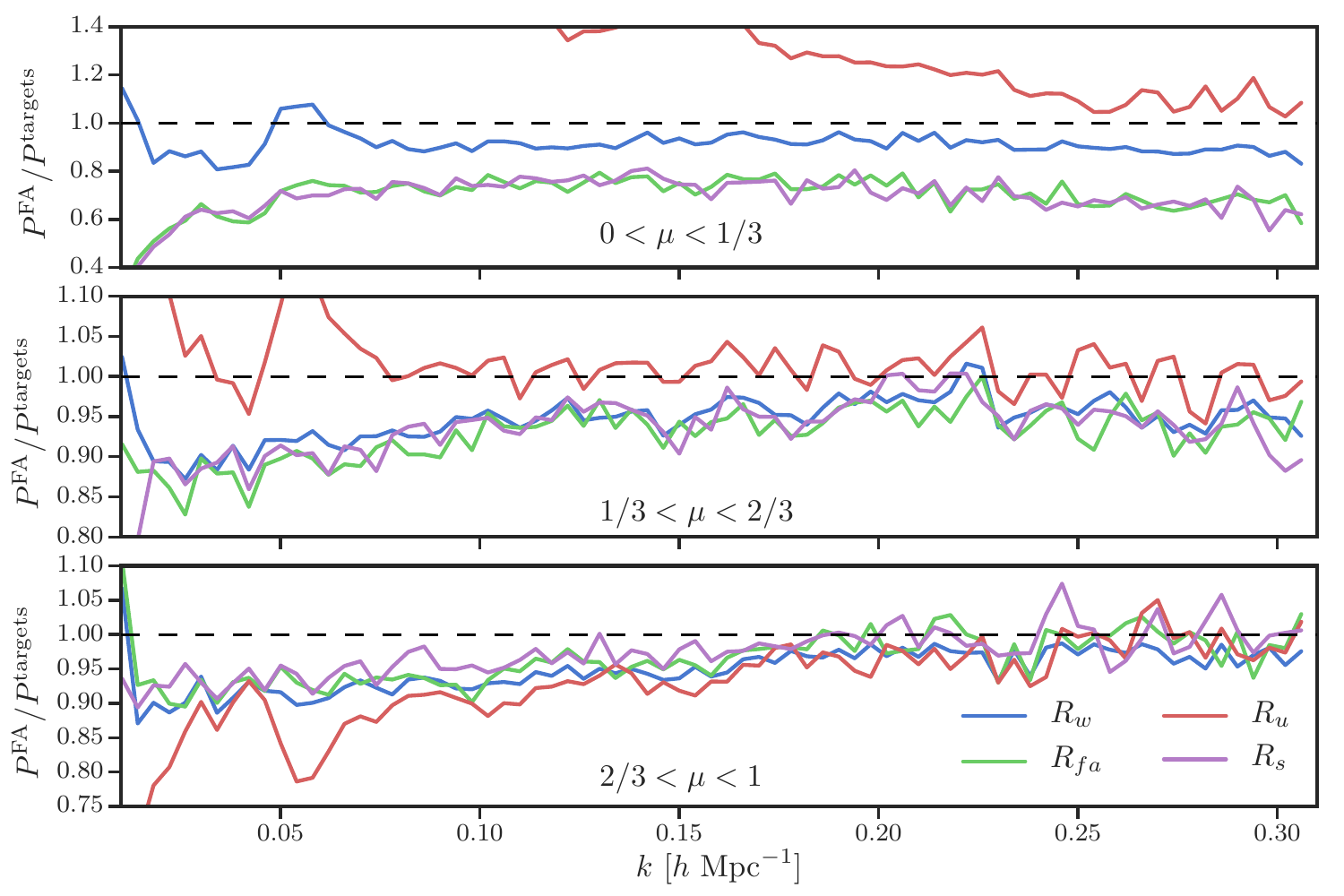}
\caption{DESI pass 1 results: the ratios of 3 $P(k,\mu)$ wedges computed 
	from the $\ell=0,2,4$ multipoles after fiber assignment to the same quantities 
    measured from the targets catalog. We compare the 
    ratios for each of the definitions of randoms described in 
    section~\ref{sec:randoms-defs}. We find the best results
    using the coverage-weighted randoms, $R_w$. }\label{fig:pass1pkmu}
\end{figure}

We compute the power spectra calculated for the four kinds of randoms presented in section \ref{sec:iso}. The multipoles are shown in figure \ref{fig:pass1poles} and the $P(k,\mu)$ analysis is shown in figure \ref{fig:pass1pkmu}. Again, \texttt{randoms\_weighted} appears 
to be the overall best choice, even though the effects are never negligible. 
The mean error is less than 5\% on the monopole, 10\% on the quadrupole, and 40\% on the hexadecapole, which is comparable to the results with the full DESI survey. The $\mu$ bins also give reasonable results, with most of the effect 
again in the lowest $\mu$ bin.

\section{Periodic Box Analysis}
\label{sec:box}

To understand better the effects of the fiber assignment, we 
perform further analysis on a periodic box.
We want to investigate the fiber assignment on a narrow $\mu$ binning, which is
simplest to analyze when using a periodic box. In this case, we have a constant number density and constant line-of-sight (the box is supposed to be seen from infinitely far away), which simplifies the analysis. We are then able to compute the power spectrum for a very large number of $\mu$ bins. The periodic box was made from a QPM simulation 
\cite{White2014}.
The analysis can be done in both redshift and real space, depending 
on whether one includes or excludes velocity information. The box is a cube of length 5.12 Gpc/$h$ and contains 143 million targets. We define $Z$ to be the line-of-sight and add velocities along this axis to get the version in redshift space of the periodic box.

Unfortunately, the DESI fiber assignment algorithm is very dependent on the survey, which cannot be easily mimicked by a periodic box simulation. 
In this section, we present some toy models of the fiber assignment and look at their effects on the power spectrum. 

\subsection{Uniform mask}

One of the effects of the fiber assignment is missing galaxies in very high-density regions. The higher the density, the more galaxies it misses. To model these non-linear effects, we first assume that the fiber assignment misses galaxies above a certain surface number density, constant across the survey.

Recall that $Z$ is the line-of-sight in our periodic box. The $X-Y$ plane is divided into $6320^2$ square cells, each one of length 0.81 Mpc/$h$. This gives a mean density of 3.58 targets per cell, equal to the average density of ELG per patrol radius in DESI. When we apply the uniform mask, we model fiber assignment by capping densities above four galaxies per cell. 

\begin{figure}[bt]
\centering    
	\begin{subfigure}[b]{0.45\textwidth}
    	\includegraphics[width=\textwidth]{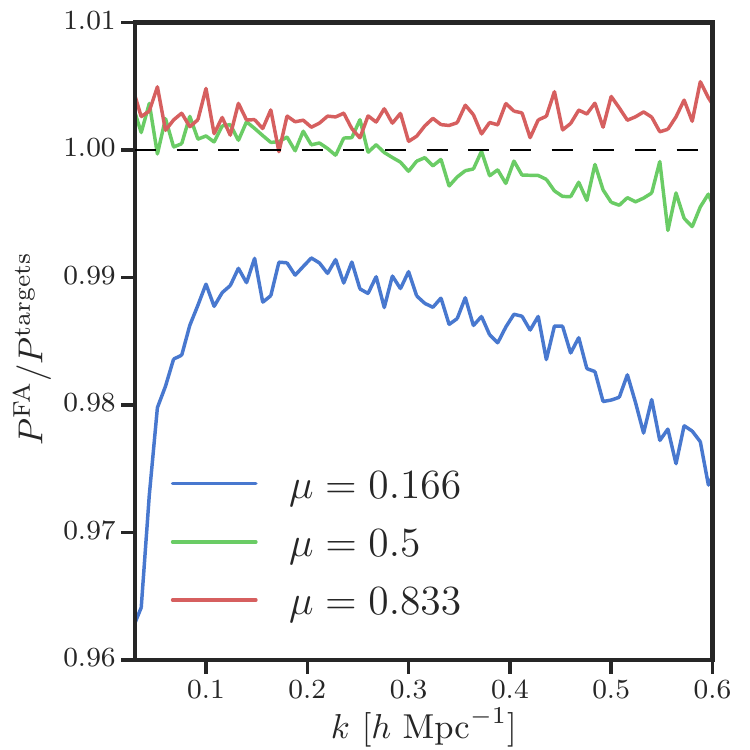}
	\caption{Redshift space}\label{fig:periodicboxpkmu3redshift}
    \end{subfigure}
	\begin{subfigure}[b]{0.45\textwidth}
    	\includegraphics[width=\textwidth]{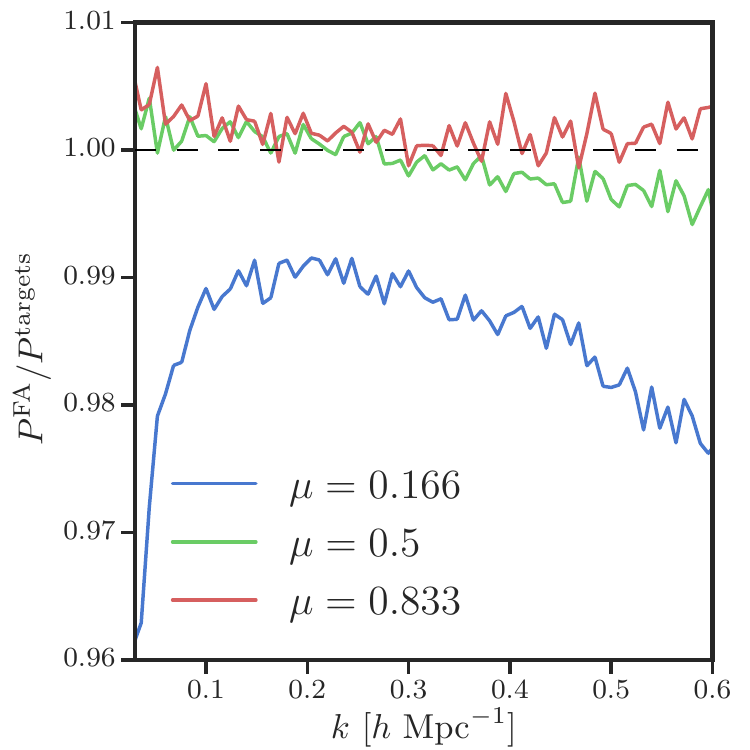}
	\caption{Real space}\label{fig:periodicboxpkmu3real}
  	\end{subfigure}
  \caption{The ratios of periodic $P(k,\mu)$ after applying a uniform fiber assignment 
  toy model divided by targets, for each of the three $\mu$ bins.}
  \label{fig:periodicboxpkmu3}
\end{figure}

\begin{figure}[bt]
\centering    
	\begin{subfigure}[b]{0.45\textwidth}
    	\includegraphics[width=\textwidth]{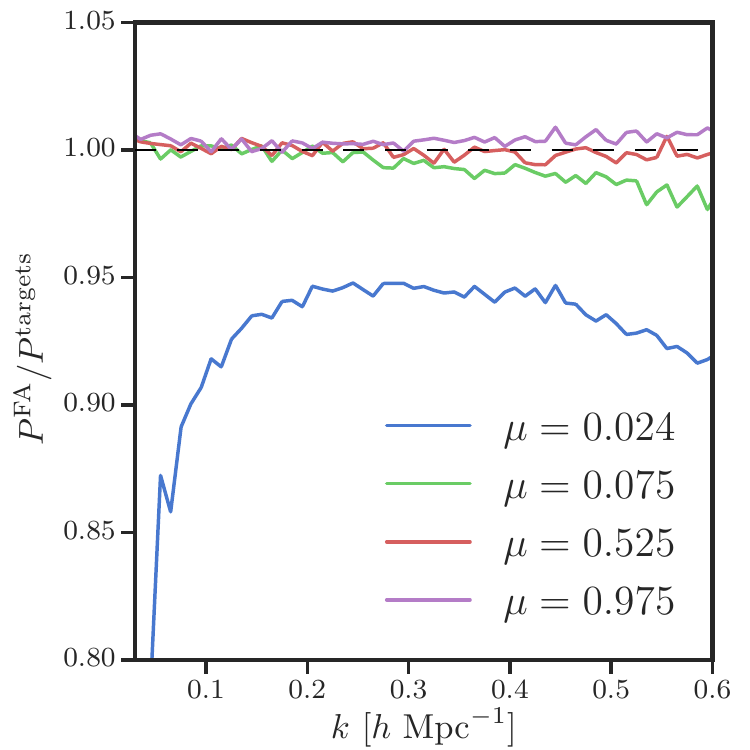}
	\caption{Redshift space}\label{fig:periodicboxpkmu20redshift}
    \end{subfigure}
	\begin{subfigure}[b]{0.45\textwidth}
    	\includegraphics[width=\textwidth]{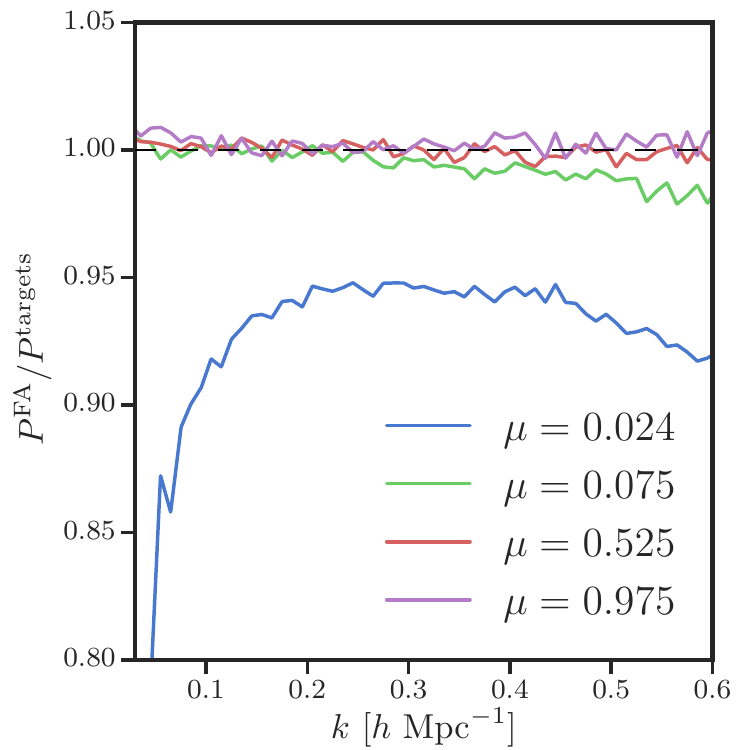}
	\caption{Real space}\label{fig:periodicboxpkmu20real}
  	\end{subfigure}
 \caption{The ratios of periodic $P(k,\mu)$ after applying a uniform fiber assignment 
 	toy model divided by targets, for 4 of the 20 $\mu$ bins.}
    \label{fig:periodicboxpkmu20}
\end{figure}

This was found to give a 16.8\% rate of rejection, close to the 21.8\% rate 
in a realistic DESI fiber assignment simulation. Because everything is uniform within 
this model, we do not need randoms to compute the power spectrum, and we can use
the \texttt{nbodykit} package to compute $P(k,\mu)$ from the periodic box before and after 
the modeled fiber assignment. The results are presented for three wide $\mu$ bins in figure \ref{fig:periodicboxpkmu3} and twenty narrow ones in figure \ref{fig:periodicboxpkmu20} (only four are presented). They are shown both in redshift and in real space, where for the 
latter we have switched off velocities in the construction of 
the galaxy catalog, making the true power spectrum isotropic (independent of $\mu$).

Once again, we observe the loss of power for the lowest $\mu$ bin, the effect being larger for low $k$. The narrower the $\mu$ bins, the larger the effect. Indeed, for the three $\mu$ bins, there is a $2\%$ effect for the first bin while for twenty bins, it is $8\%$ for the first bin. Already for the second of the 20 bins, $\mu \sim 0.075$, the effect is at a sub-percent level. This strongly supports the idea that only the $\mu$=0 modes, i.e. in the plane perpendicular to the line-of-sight, are affected by the fiber assignment. We observe that the effects are similar in redshift and real space, where for 
the latter we do not add peculiar velocities to the galaxies (hence no multipole with $l>0$ is generated). 

\subsection{Spatially varying mask}

A more realistic model of the fiber assignment allows the coverage, i.e. the number of observed galaxies in a patrol 
area, to vary across the survey. 
We choose a two-dimensional sinusoidal mask.
Recalling that the $(x,y)$ plane is divided into $6300^2$ cells of 0.81 Mpc/$h$ of length, indexed by $(j,k)$, the functional form is

\begin{equation}
\label{eqn:sinumask}
W_{jk}=4+ \mathrm{sin}\left[(\lfloor j/40 \rfloor + \lfloor k/40 \rfloor) \frac{\pi}{2}\right]
\end{equation}
where $\lfloor . \rfloor$ is the ceiling function (the smallest integer greater than the argument), and in practice, $W_{jk}\in \lbrace 3,4,5 \rbrace$, and its value is twice more often 4 than 3 or 5. The mask is constant over squares of 40$^2$ cells and has a wavelength of 130 Mpc/$h$. We find that this new model rejects 18.1\% of the targets.

\begin{figure}[bt]
\centering    
    \includegraphics[width=0.8\textwidth]{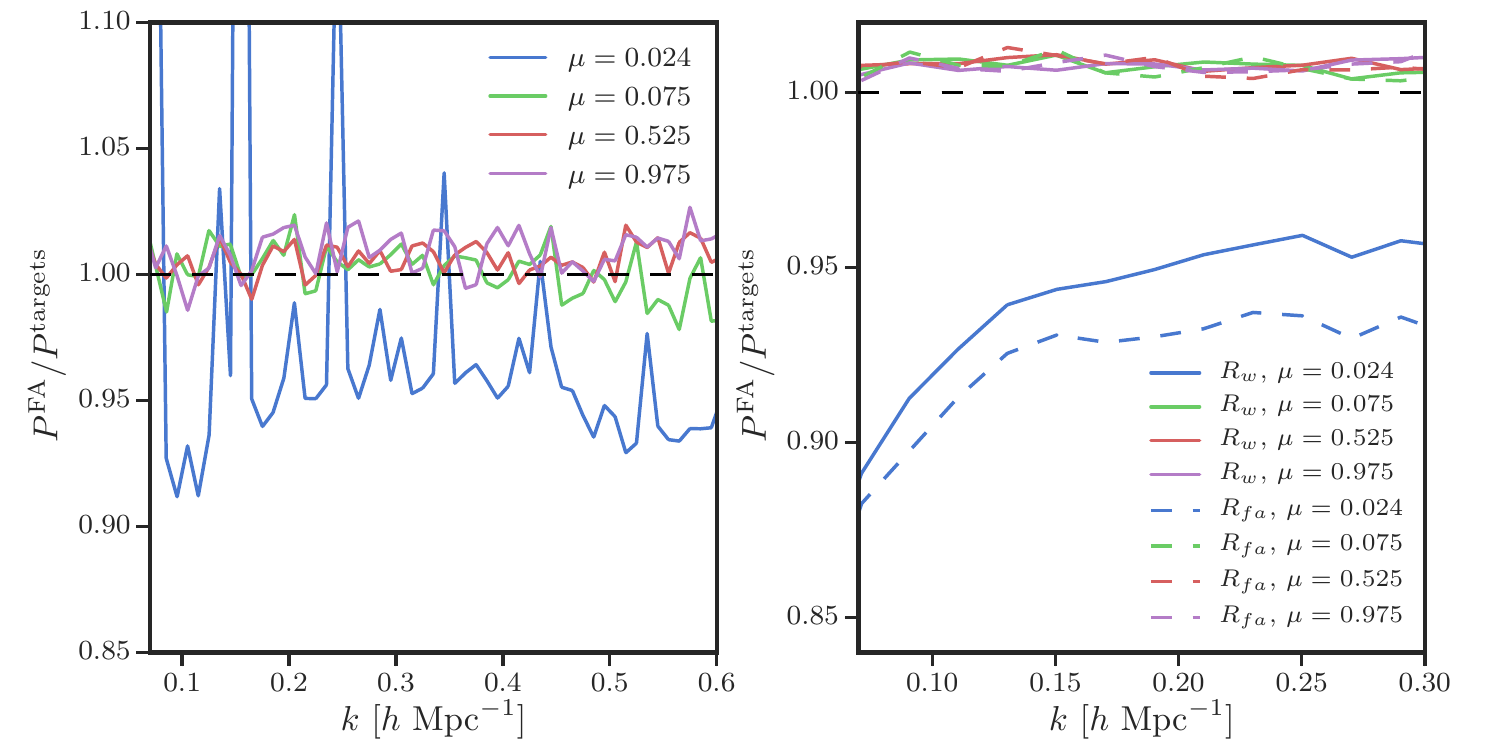}
 \caption{The ratios of periodic $P(k,\mu)$ after applying a spatially varying fiber
 	assignment toy model divided by targets, for 4 of the 20 $\mu$ bins. The results
    are computed in real space, where all $\mu$ bins have equal power.}
    \label{fig:varyingperiodicbox}
\end{figure}

For the spatially varying mask 
we need to use the randoms presented previously. In the left panel of 
figure \ref{fig:varyingperiodicbox}, we show the naive power spectrum obtained with 
\texttt{randoms\_uniform}, and we see the large correlations induced for $\mu \sim 0$ modes. 
But as one can see in the right panel of figure \ref{fig:varyingperiodicbox}, using 
\texttt{randoms\_after\_fa} and \texttt{randoms\_weighted} enables one to recover the 
power spectrum, with a 7\% and 5\% offsets for $\mu \sim 0$, respectively, 
and sub-percent errors for $\mu>0$ bins. 

This periodic box analysis provides support for the idea that 
most of the contaminating signal introduced by fiber assignment is in angular ($\mu=0$) modes, and that using the appropriate randoms enables the recovery of the true power spectrum for all modes that are not perpendicular to the line-of-sight. 

\begin{figure}[bt]
\begin{center}
\includegraphics[width=0.7\textwidth]{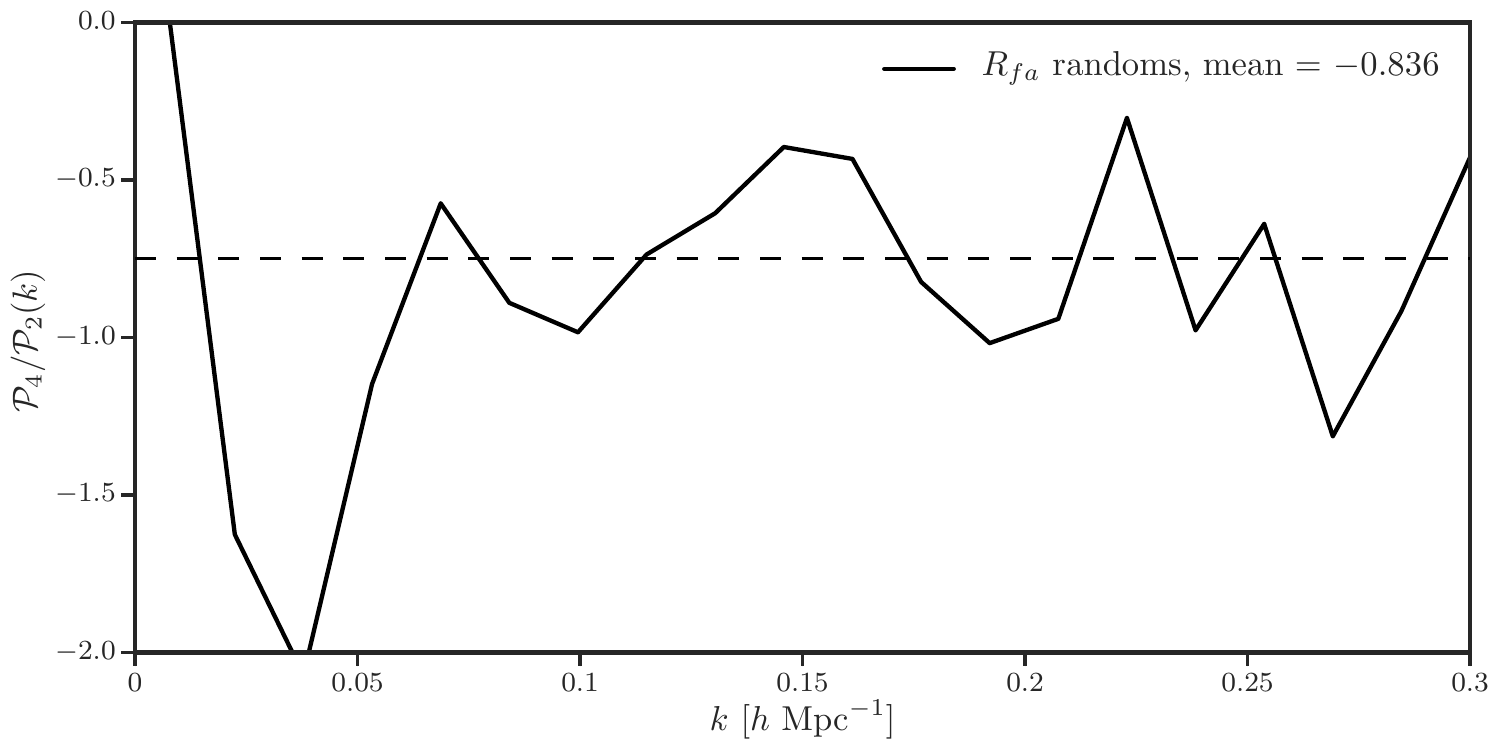}
\caption{The ratio $\mathcal{P}_4 (k) / \mathcal{P}_2 (k)$ for DESI after fiber assignment, 
	in real space, when using \texttt{randoms\_after\_fa} randoms. The expected ratio is
    constant and equal to $-3/4$, if the only contaminated mode is $\mu=0$.}
    \label{fig:hexaoverquad}
\end{center}
\end{figure}

We have seen that the analysis is simplest for $P(k,\mu)$, 
which poses an analysis challenge for a full survey with its 
geometry, where the analysis has typically been done with 
multipoles. 
In our analysis of $\mu$ bins, we have used the first three multipoles 
and thus assumed that multipoles with $\ell>4$ are zero, an assumption that is explicitly 
violated for the case where we have significant 
transverse power at $\mu=0$. We have tested our 
bin reconstruction by using finely sampled 
periodic box $P(k,\mu)$, 
computing the multipoles, and then using equation \ref{eqn:pkmu-from-poles}
to compute three broad bins, and compare the results to the direct 
computation of three bins, finding of order 1\% 
differences. 
We can improve upon this method by assuming a smooth 
model for the true power spectrum, plus an additional 
component at $\mu=0$, which we call $P_c$. 
The simplest case is when the higher order 
multipoles are absent, such as is the case for $\ell>4$ on large scales. Here, 
we will look at the even simpler case 
of real space clustering, where all multipoles with $l>0$ are zero. 
In this case, the $l>0$ multipoles after 
fiber assignment are given by 
$\mathcal{P}_\ell = P_c \mathcal{L}_\ell(\mu=0)$,
where $\mathcal{L}_\ell(\mu)$ is the Legendre polynomial of order $\ell$. In figure \ref{fig:hexaoverquad}, we plot the ratio $\mathcal{P}_4 (k) / \mathcal{P}_2 (k)$ of DESI after fiber assignment, in real space. Under previous assumption, the ratio should thus be constant and equal to $-3/4$. 
We see that this is in a good agreement with the simulations. 
More generally, 
by modeling the component that is smooth in $\mu$ and $P_c$ at the same time one can separate the two using the multipole
analysis. This technique should be developed further combining it with state-of-the-art RSD modeling.

\section{Conclusions}\label{sec:conclude}

In this paper, we investigate the effects of the DESI survey strategy, including tiling and the fiber assignment algorithm, on the power spectrum of emission line galaxies. We 
investigate several methods of defining the mean density of the survey (or, 
equivalently, randoms catalogs), that have different effects on 
the clustering. We find that a uniform randoms density which 
includes no fiber assignment effects has the worst performance, because it 
does not include the spatial variation of the mean density introduced by fiber assignment.
In contrast, 
computing the mean density separately  according to the value of the coverage at the position of each galaxy
reduces the effects of fiber assignment and projects most of the 
residual effect on to the lowest $\mu$-bin, i.e. the plane perpendicular to the line-of-sight.
For the nominal DESI five-year survey, the remaining 
effects are of order a few percent, 
which one should be able to calibrate out using realistic 
simulations. The effects after one year are larger, of order 
10\%, and it remains to be seen if they can be calibrated 
out. From these results and the DESI forecasts of \cite{DESIScienceDesign:2016}, 
we expect systematics related to fiber assignment to be able to be controlled 
for BAO measurements, where a specific feature is being isolated independent
of the broadband clustering. RSD measurements on individual redshift bins are
forecasted to yield no more than percent level constraints of $f\sigma_8$,
and we are optimistic that the level of systematics can be sufficiently controlled. 
Further analysis of calibration techniques for residual systematics must be carefully
studied, especially in the context of constraints from full-shape, 
broadband clustering information, i.e., the sum of the 
neutrino masses.

We have also explored two other types of 
randoms, which are less effective, but perform better than 
uniform randoms, and also show most 
of the fiber assignment contamination to be localized in the 
transverse $\mu =0$ bin. 
We confirm these results using 
a periodic box analysis, where one is able to 
compute the power spectrum using narrow $\mu$ bins. 
We then repeat the analysis using a mock simulation of fiber assignment 
on a true periodic box, inducing the modulation purely 
in $X-Y$ plane. Even in 
the case of very narrow bins, we still see the dominant effect 
to be localized in $\mu=0$: this is not surprising, since by 
construction there is no modulation in the $Z$ direction, 
implying that only the $k_Z=0$ mode is affected. 
We emphasize that the effects can only be localized (to $\mu=0$) in a three-dimensional power spectrum analysis, while the 
corresponding three-dimensional correlation function analysis does not 
localize these effects.

In addition to obtaining realistic simulations 
for calibration of residual effects, 
the main remaining issue is the lack of an estimator to 
efficiently perform a $P(k,\mu)$ analysis with narrow 
$\mu$ bins for a realistic survey. 
Current state-of-the-art Fourier space analyses \cite{Grieb2016,Beutler2016,Gil-Marin2016} use a FFT-based
multipole analysis to measure $\mathcal{P}_\ell(k)$. The number of 
FFTs, and hence computational cost, 
increases rapidly with $\ell$ \cite{Bianchi2015, Scoccimarro2015}. 
Multipoles can be used to calculate binned values $P(k,\mu)$ \cite{Grieb2016}, where this calculation assumes 
higher-order ($\ell>4$) multipoles are zero. 
One can always compute 
higher-order multipoles to test this assumption and 
improve the calculation, at a cost of increased complexity 
and number of FFTs \cite{Bianchi2015, Scoccimarro2015}. To date, no analysis has gone beyond 
$\ell=4$ because one does not expect there to be much signal at $\ell>4$, but it should be possible to increase this to a higher 
$\ell$, if needed, to remove the transverse mode systematics. 
If one decides to simply remove the lowest $\mu$ bin the 
error increases roughly by $1/(2M)$, where $M$ is the number of the bins, 
so with $M=5$ (corresponding to analysis up to $\ell=10$)
we expect the statistical error to increase by 10\%. 
In our tests, we find that when using weighted randoms, the effects
are a few percent even for the lowest $\mu$ bin, 
(except for very low $k$, where the effect can be as high 
as 10\%), so when using these randoms 
it may be possible to calibrate out the effects 
without removing low $\mu$ bins, and hence continue to use the
analysis up to $\ell=4$ only. 

An alternative is to employ a full spherical basis, 
where the modes are already in the basis suitable for 
removal of purely angular modes \cite{Heavens1995,Percival2004}. One cannot 
employ FFTs in this basis and the overall scaling is slower
(from $O(N\log N)$ to $O(N^{4/3}$)), 
but this is compensated by the need to have only one transform instead of $21$ for $\ell=0,2,4$ analysis. 
The main 
deficiency of this basis is that it does not localize 
a given scale in comoving coordinates, and hence does 
not localize BAO features. It remains to be seen if 
the benefits of this basis outweigh its deficiencies. 

\acknowledgments{
We thank Pat McDonald and Nikhil Padmanabhan for useful discussions and
Julien Guy for comments on the manuscript. 
This research is supported by the Director, Office of Science, Office of 
High Energy Physics of the U.S. Department of Energy under Contract No. 
DE–AC02–05CH1123, and by the National Energy Research Scientific Computing Center, 
a DOE Office of Science User Facility under the same contract; additional support 
for DESI is provided by the U.S. National Science Foundation, Division of 
Astronomical Sciences under Contract No. AST-0950945 to the National Optical 
Astronomy Observatory; the Science and Technologies Facilities Council of the United
Kingdom; the Gordon and Betty Moore Foundation; the Heising-Simons Foundation; 
the National Council of Science and Technology of Mexico, and by the DESI 
Member Institutions. The authors are honored to be permitted to conduct astronomical
research on Iolkam Du’ag (Kitt Peak), a mountain with particular significance to the 
Tohono O’odham Nation. 
NH is supported by the National Science Foundation Graduate Research Fellowship under 
grant number DGE-1106400. US is supported by NASA grant NNX15AL17G. 
}

\bibliographystyle{JHEP}
\bibliography{refs}
\end{document}